\documentclass[preprint,nofootinbib,showkeys]{revtex4}

\usepackage{graphicx}
\usepackage{multirow,array}
\usepackage{amssymb}
\usepackage{setspace}
\usepackage{amsmath}
\usepackage{textcomp}

\newcommand{\bee}{\begin{equation}}
\newcommand{\eee}{\end{equation}}
\newcommand{\eaa}{\end{eqnarray}}
\newcommand{\baa}{\begin{eqnarray}}

\newcommand{\p}{\partial}
\newcommand{\al}{\alpha}
\newcommand{\be}{\beta}

\newcommand{\ta}{\theta}
\newcommand{\de}{\delta}
\newcommand{\ga}{\gamma}

\newcommand{\g}{\Gamma}
\newcommand{\om}{\omega}

\def\ni{\noindent}

\begin{document}

\title{\large{Noncommutative analysis in a curved phase-space \\ and coherent states quantization}}

\author{B. F. Rizzuti$^{a}$}
\email{brunorizzuti@ufam.edu.br}

\author{E. M. C. Abreu$^{b,c}$}
\email{evertonabreu@ufrrj.br}

\author{A. C. R. Mendes$^{c}$}
\email{albert@fisica.ufjf.br}

\author{M. A. Freitas$^a$}

\author{V. Nikoofard$^c$}
\email{vahid@fisica.ufjf.br}

\affiliation{$^a$Instituto de Sa\'ude e
Biotecnologia, ISB, Universidade Federal do Amazonas, 69.460-000, Coari, AM, Brazil}

\affiliation{$^b$Grupo de F\' isica Te\'orica e Matem\'atica F\' isica, Departamento de F\'{i}sica, Universidade Federal Rural do Rio de Janeiro, 23890-971, Serop\'edica, RJ, Brazil}

\affiliation{$^c$Departamento de F\'{i}sica, Universidade Federal de Juiz de Fora, 36036-330, Juiz de Fora, MG, Brazil}

\begin{abstract}
\ni In this work we have shown precisely that the curvature of a 2-sphere introduces quantum features in the system through the introduction of 
the noncommutative (NC) parameter that appeared naturally via equations of motion.
To obtain this result we used the fact that quantum mechanics can be understood as a NC symplectic geometry, which generalized the standard description of classical mechanics as a symplectic geometry.  
In this work, we have also analyzed the dynamics of the model of a free particle over a 2-sphere in a NC phase-space. 
Besides, we have shown the solution of the equations of motion allows one to show the equivalence between the movement of the particle physical degrees of freedom upon a 2-sphere and the one described by a central field. We have considered the effective force felt by the particle as being caused by the curvature of the space.  We have analyzed the NC Poisson algebra of classical observables in order to obtain the NC corrections to Newton's second law.   We have demonstrated precisely that the curvature of the space acted as an effective potential for a free particle in a flat phase-space.  Besides, through NC coherent states quantization we have obtained the Green function of the theory.  The result have confirmed that we have an UV cutoff for large momenta in the NC kernel.
 We have also discussed the relation between affine connection and Dirac brackets, as they describe the proper evolution of the model over the surface of constraints in the Lagrangian and Hamiltonian formalisms, respectively. As an application, we have treated the so-called \textit{Zitterbewegung} of the Dirac electron. Since it is assumed to be an observable effect, then we have traced its physical origin by assuming that the electron has an internal structure. 
\end{abstract}

\date{\today}


\maketitle

\section{Introduction}

The great interest in noncommutative (NC) space-time theories nowadays had inspired the analysis of both several models/theories and the behavior of the divergences in this ``new" regime.   The last one, by the way, was the original motivation that made Snyder to publish the first paper on the subject \cite{snyder}.   However, since it was demonstrated that the tame of the infinities of QFT  was not accomplished \cite{yang}, the NC formalism was put to sleep for more than fifty years.   The string formalism \cite{sw} has awaked the noncommutativity procedure and since then we have witnessed the growing interest on noncommutativity \cite{reviews}.   One of these interests is to observe what is the contribution of this formalism concerning both classical mechanics and general relativity.   Since the introduction of noncommutativity means (in general) the introduction of the so-called NC parameter, which has area dimension and its value is of Planck scale, we can say that, when introduced in classical theories, it could mean a kind of semi-classical approach.   Therefore, considering general relativity, noncommutativity is one candidate to obtain a path to quantize gravitation, for instance.   Considering other classical systems, the introduction of noncommutativity can be realized as a link to Planck scales, like quantum mechanics and its possible relation to UV/IR mixing \cite{djemai}.   

In the specific case of classical mechanics, considered here, we can analyze the contribution of noncommutativty in order to add a perturbation in Newton's second law for the systems considered \cite{rsv}.   Namely, since the equations of motions are modified, when treated in a NC space, we can ask about the effects in the acceleration coordinate \cite{rsv, djemai}.   It is one of the targets of this paper, together with the divergence analysis through coherent states quantization.  But first, we will describe classically our system, a free particle in a curved space, i.e., a 2-sphere.

Classical mechanics is one of the most enlightening starting points for introducing many distinct mathematical tools such as differential equations, symplectic structures \cite{arno}, and in particular, the basic concepts of differential geometry. For example, in \cite{aadeurop}, the author used a potential motion to construct the corresponding geometric setting. In this way, some notions such as Riemann metric space, Christoffel symbols, parallel transport and covariant derivative are introduced. We extend this idea here. Instead of treating a potential motion, we will describe a free particle constrained to a curved surface. By constructing its corresponding Lagrangian, we are naturally led to a free motion in a Riemann space. Definitions of metric and Christoffel symbols appear in the course of constructing the dynamics of the model. We will analyze in details the movement of a particle over a 2-sphere. 
Solution of the equations of motion are given in two different ways. Firstly we will explore the geometrical properties of the model and after that, we will use 
Noether charges to decouple equations of motion. Moreover, due to the symmetrical structure of the 2-sphere, we will establish the equivalence between the motion in a central field and the free particle over the 2-sphere. It turns out that the central potential is proportional to the curvature of the surface. Then, constrained systems may be also a suitable analogue formalism to introduce general relativity, once Einstein interpreted gravity as a deformation of space-time due to the presence of mass \cite{gr}. We will also treat the corresponding hamiltonization of the free particle over the 2-sphere according to the Dirac algorithm for constrained systems \cite{dirac}, which enables one to establish the intrinsic relation between the Dirac brackets and Christoffel symbols, since both of them are supposed to provide the proper evolution over the surface where the model is defined, the first in the phase space and the former, in the configuration space. Although all the calculations are performed classically, we discuss an application in the quantum realm. 
We set one possible interpretation of the so-called \textit{Zitterbewegung}, a quivering motion predicted by Schr\"odinger when he scrutinized the Dirac equation \cite{sch}. The time evolution of electron position operators may be separated in two parts: one in a rectilinear movement and the other oscillates in a ellipse as trajectory, resembling the physical variables of a free particle over a 2-sphere.  Thus, the  \textit{Zitterbewegung} may be interpreted as a position variable constrained to a 2-sphere if we assume an internal structure to the electron.

The paper is organized as follows. In Section 2, we will discuss an alternative way to introduce a constraint into a Lagrangian. We show the equivalence between the formulation of describing the model in terms of physical variables and the one where the constraint is inserted via Lagrange multipliers. Sections 3 and 4 will be dedicated to a detailed analysis of a particle over a 2-sphere. The construction of the action in terms of physical variables and its limits according to the principle of least action lead naturally to the concepts of metric and affine connection. We will also obtain the general solution of the equations of motion. 
In Section 5 we will establish the equivalence between the movement in a central field and the one taken by the physical variables of our particle over a 2-sphere. 
In Section 6 we will introduce noncommutativity through the equations of motion via symplectic framework and Poisson brackets.   We will demonstrate that the space curvature introduced naturally the NC parameter, which can be understood as the introduction of quantum features in the classical system, or at least of a semi-classical feature.
 The Section 7 will explain the NC coherent states formalism and the kernel and Green functions will be obtained.
In  Section 8 we will provide the hamiltonization of the constrained system described in the previous Sections. The application concerning the electron \textit{Zitterbewegung} will be discussed in Section 9. Section 10 will be dedicated to the conclusions and perspectives.               

\section{Constrained systems: the basic formalism}

The basic directions to introduce a constraint into a Lagrangian is via Lagrange multipliers. Equivalently, knowing \textit{a priori} the constraint of the model, one may find one of the variables in terms of the others and include it into the Lagrangian, leading to a new formulation in terms of physical variables, \textit{i.e.}, whose dynamics is independent of the remaining ones. Our first step in this notes is to show the equivalence between the new and former formulations. Besides, this Section also fixes the notation which shall be used throughout the paper. 

Let us consider a free particle constrained to the surface
\begin{eqnarray}\label{01}
\Phi(x^i)=0,
\end{eqnarray}
where $x^i = x^i(t)$; $i=1,...,N$ are the coordinates of the system. Under technical conditions satisfied by the function $\Phi$, we may find one of the variables, say $x^1$, in terms of the others,
\begin{eqnarray}\label{02}
\Phi(x^i)=0 \Leftrightarrow x^1 = f(x^{\al}); \, \al=2,...,N.
\end{eqnarray}

\ni From now on in this section, Greek letters mean the values $2,...,N$. In this case, $x^1$ is a non-physical degree of freedom because its dynamics is dependent of the remaining variables $x^{\al}$. If $L = L(x^i, \dot x^i)$ is the Lagrangian of the free particle in the absence of the constraint (\ref{01}), then the prescription to construct an action in terms of the physical variables $x^{\al}$ is the following,
\begin{eqnarray}\label{03}
S_1 = \int dt L(x^i, \dot x^i)|_{x^1 = f(x^{\al})},
\end{eqnarray}
where we have denoted $\dot x^i \equiv \frac{dx^i}{dt}$. With more details,
\begin{eqnarray}\label{04}
L(x^i, \dot x^i)|_{x^1 = f(x^{\al})} = L(x^1=f(x^\al), \dot x^1= \frac{\p f}{\p x^\be} \dot x^\be, x^\al, \dot x^\al) \equiv \bar L (x^\al, \dot x^\al).
\end{eqnarray}

\ni The notation $\bar L$ indicates the substitution of $x^1=f(x^\al)$ in (\ref{03}) and repeated indexes mean summation, as usual. To obtain the Euler-Lagrange equations of (\ref{03}), we evaluate separately the derivatives of the expression (\ref{04}),
\begin{eqnarray}\label{05}
\frac{\p \bar L(x^\al, \dot x^\al)}{\p x^\ga}=\frac{\p L (x^i, \dot x^i)}{\p x^1} \Big |\frac{\p f}{\p x^\ga} + \frac{\p L(x^i, \dot x^i)}{\p \dot x^1} \Big | \frac{\p^2 f}{\p x^\ga \p x^\be}\dot x^\be + \frac{\p L(x^i, \dot x^i)}{\p x^\ga} \Big |, 
\end{eqnarray}
\begin{eqnarray}\label{06}
\frac{\p \bar L (x^\al, \dot x^\al)}{\p \dot x^\ga}= \frac{\p L(x^i, \dot x^i)}{\p \dot x^1} \Big | \frac{\p f}{\p x^\ga} + \frac{\p L (x^i, \dot x^i)}{\p \dot x^\ga} \Big |,
\end{eqnarray}

\ni where $|$ corresponds to the substitution expressed in (\ref{03}). It will be used in subsequent calculations. Hence, the equations of motion given by
\begin{eqnarray}\label{07}
\frac{\de S_1}{\de x^\ga} = \frac{\p \bar L(x^\al, \dot x^\al)}{\p x^\ga}-\frac{d}{dt}\left (\frac{\p \bar L(x^\al, \dot x^\al)}{\p \dot x^\ga} \right)=0
\end{eqnarray}
provides, after rearranging the terms,
\begin{eqnarray}\label{08}
\frac{\de S_1}{\de x^\ga}\Big | + \frac{\de S_1}{\de x^1}\Big | \frac{\p f}{\p x^\ga}=0.
\end{eqnarray}

The idea here is to show that one may insert the constraint $\Phi(x^i)=0$ into the initial Lagrangian leading to an equivalent description. Let us consider the following action,
\begin{eqnarray}\label{09}
S_2 = \int dt \tilde L (x^i, \dot x^i, \lambda),
\end{eqnarray}

\ni defined in an extended configuration space parametrized by $x^i$, $\lambda$, where
\begin{eqnarray}\label{10}
\tilde L (x^i, \dot x^i, \lambda) = L(x^i, \dot x^i)+ \lambda \Phi(x^i).
\end{eqnarray}
The functions $L$ and $\Phi$ are the same as the initial construction and $\lambda$ is a Lagrange multiplier. Hence, the Euler-Lagrange equations are
\begin{eqnarray}\label{11}
\frac{\de S_2}{\de x^1} = 0 \Rightarrow \frac{\p L(x^i, \dot x^i)}{\p x^1} + \lambda \frac{\p \Phi}{\p x^1} = \frac{d}{dt}\left (\frac{\p L(x^i, \dot x^i)}{\p \dot x^1} \right),
\end{eqnarray}
\begin{eqnarray}\label{12}
\frac{\de S_2}{\de x^\ga} = 0 \Rightarrow \frac{\p L(x^i, \dot x^i)}{\p x^\ga} + \lambda \frac{\p \Phi}{\p x^\ga} = \frac{d}{dt}\left (\frac{\p L(x^i, \dot x^i)}{\p \dot x^\ga} \right),
\end{eqnarray}
\begin{eqnarray}\label{13}
\frac{\de S_2}{\de \lambda} = 0 \Rightarrow \Phi(x^i) = 0.
\end{eqnarray}
From (\ref{11}), we find
\begin{eqnarray}\label{14}
\lambda = -\left (\frac{\p \Phi}{\p x^1} \right )^{-1}\left [\frac{\p L(x^i, \dot x^i)}{\p x^1}-\frac{d}{dt}\left (\frac{\p L(x^i, \dot x^i)}{\p \dot x^i} \right) \right ].
\end{eqnarray}

The substitution of (\ref{14}) in (\ref{12}) eliminates the $\lambda$-dependence of equations of motion,
\begin{eqnarray}\label{15}
\frac{\p L(x^i, \dot x^i)}{\p x^\ga}-\frac{d}{dt}\left (\frac{\p L(x^i, \dot x^i)}{\p \dot x^\ga} \right) - \left (\frac{\p \Phi}{\p x^1} \right )^{-1}\frac{\p \Phi}{\p x^\ga}\left [\frac{\p L(x^i, \dot x^i)}{\p x^1} -\frac{d}{dt}\left (\frac{\p L(x^i, \dot x^i)}{\p \dot x^1} \right ) \right ]=0.
\end{eqnarray}
Finally, from (\ref{13}) and according to (\ref{02}),
\begin{eqnarray}\label{16}
\Phi(x^i) = 0 \Leftrightarrow x^1= f(x^\al).
\end{eqnarray}
Substitution of $x^1= f(x^\al)$ into the constraint $\Phi(x^i)=0$ gives the identity
\begin{eqnarray}\label{17}
\Phi(x^1= f(x^\al), x^\al) \equiv 0,
\end{eqnarray}
whose derivative provides
\begin{eqnarray}\label{18}
0=\frac{d}{dx^\ga}\Phi(x^1= f(x^\al), x^\al) = \frac{\p \Phi(x^i)}{\p x^1}\Big | \frac{\p f}{\p x^\ga}+ \frac{\p \Phi(x^i)}{\p x^\ga}\Big |.
\end{eqnarray}
Then we have that
\begin{eqnarray}\label{19}
\frac{\p f}{\p x^\ga} = -\left[\frac{\p \Phi(x^i)}{\p x^1}\Big | \right]^{-1}\frac{\p \Phi(x^1)}{\p x^\ga}\Big |.
\end{eqnarray}
This expression appears in (\ref{15}), which is now rewritten by eliminating $x^1$,
\begin{eqnarray}\label{20}
\left[\frac{\p L(x^i, \dot x^i)}{\p x^\ga}-\frac{d}{dt}\left(\frac{\p L(x^i, \dot x^i)}{\p \dot x^\ga} \right) \right]\Big |+ \left[\frac{\p L(x^i, \dot x^i)}{\p x^1}-\frac{d}{dt}\left(\frac{\p L(x^i, \dot x^i)}{\p \dot x^1} \right) \right]\Big | \frac{\p f}{\p x^\ga}=0. 
\end{eqnarray}
Since $\frac{d}{dt}\left(L|\right)=\frac{dL}{dt} |$, we arrive at
\begin{eqnarray}\label{21}
\frac{\de S_1}{\de x^\ga}\Big | + \frac{\de S_1}{\de x^1}\Big | \frac{\p f}{\p x^\ga}=0.
\end{eqnarray}

\ni These are the same equations of motion of the initial formulation, see (\ref{08}). The equivalence between both constructions that have been developed so far becomes clearer if we compare the number of degrees of freedom in each description. The initial construction described by $\bar L = \bar L(x^\ga, \dot x^\ga)$ was formulated by eliminating $x^1$ with the previous knowledge of the constraint surface the model is immersed in. We are left $N-1$ degrees of freedom. On the other hand, the second one starts with $N+1$ variables. First, we have excluded $\lambda$ from the description by using (\ref{11}). Then, with the help of (\ref{13}), $x^1$ was eliminated, see (\ref{16}). These two steps left us with $N+1-2=N-1$ degrees of freedom, as expected. This concludes the equivalence between $S_1$ and $S_2$. An application will be treated in the next Section, when we consider the example of a particle over a 2-sphere.

\section{A concrete example of constrained dynamics: particle over a 2-sphere}

We shall now discuss an application of the result found in the last Section. Actually, the main aim of these notes is the classical and NC descriptions of a free particle over a 2-sphere. Besides,  the example of the particle over a 2-sphere will be used for a classical description of the Dirac spinning electron, see Section 9. 

Let $m$ be the mass of the particle and $x^i=x^i(t)$, $i=1,2,3$, its spatial coordinates. Since we want to formulate the particle evolution constrained to a 2-sphere, we take the following action,
\begin{eqnarray}\label{22}
S_{\lambda} = \int dt \left[\frac{m}{2}\de_{ij} \dot x^i \dot x^j+ \lambda(\de_{ij}x^i x^j-a^2) \right],
\end{eqnarray}  
where $\de_{ij}$ stands for the delta Kronecker symbol and $\lambda$ is a Lagrange multiplier. $S_{\lambda}$ has manifest $SO(3)$-invariance, which guarantees, for example, conservation of angular momentum. Equation of motion for $\lambda$ gives the desired constraint
\begin{eqnarray}\label{23}
\de_{ij}x^i x^j = a^2.
\end{eqnarray} 
So, (\ref{22}) in fact describes a free particle over a 2-sphere of radius $a$. On the other hand, we could exclude one of the variables with the help of (\ref{23}),
\begin{eqnarray}\label{24}
x^3 = \pm \sqrt{a^2 - \de_{ij}x^i x^j},
\end{eqnarray}
where $i,j$ run the values $1$ and $2$. Concerning the parametrization of the 2-sphere, we take the upper half plane $x^3>0$. Then, according to (\ref{03}), we substitute (\ref{24}) into the action for the free particle in a flat 3-dimensional space leading to 
\begin{eqnarray}\label{25}
S_{ph} = \int dt \frac{m}{2}g_{\al \be}\dot x^\al \dot x^\be,
\end{eqnarray}
where
\begin{eqnarray}\label{26}
g_{\al \be}(x) = \de_{\al \be}+ \frac{x_\al x_\be}{a^2 - \de_{ij}x^i x^j}.
\end{eqnarray}
The action was named $S_{ph}$ since we have eliminated the spurious degree of freedom $x^3$, obtaining an equivalent description of the particle over a 2-sphere in terms of physical variables $x^1$, $x^2$. It has a simple interpretation: since the particle is constrained to a 2-sphere, (\ref{25}) describes a free particle in a Riemann space whose metric is given by $g_{\al \be}$ \cite{aadbook}. The elimination of $x^3$ naturally led us to the concept of first fundamental form (or metric) \cite{manf}. In the limit $a \rightarrow + \infty$, we have a free particle in a flat bi-dimensional space. Namely, $g_{\al \be} \rightarrow \de_{\al \be}$ and the Lagrangian originated from (\ref{25}) becomes the kinetic energy of the particle,
\begin{eqnarray}\label{27}
\frac{m}{2}g_{\al \be}\dot x^\al \dot x^\be \rightarrow  \frac{m}{2}\left[(\dot x^1)^2 + (\dot x^2)^2 \right].
\end{eqnarray} 

We now turn our attention to the time evolution of the model. The dynamics is governed by the principle of least action. The minimization $\de S_{ph}=0$ gives the equation of motion
\begin{eqnarray}\label{28}
\ddot x^\al = G^\al{}_{\sigma \be}\dot x^\sigma \dot x^\be,
\end{eqnarray}
where
\begin{eqnarray}\label{29}
G^\al{}_{\sigma \be} = g^{\al \ga}\left(\frac{1}{2}\p_\ga g_{\sigma \be} - \p_\sigma g_{\ga \be} \right).
\end{eqnarray}
$g^{\al \gamma}$ corresponds to the inverse of the metric: $g^{\al \be}g_{\be \ga} = \de^\al{}_\ga$  and $\p_\ga \equiv \frac{\p}{\p x^\ga}$. Explicit calculation of $G$ gives
\begin{eqnarray}\label{30}
G^\al{}_{\sigma \be} = \frac12 \frac{x_\sigma \de^\al{}_\be - x_\be \de^\al{}_\sigma}{a^2-\de_{\ga \rho} x^\ga x^\rho} - \frac{x^\al g_{\sigma \be}}{a^2}.
\end{eqnarray}
The first term of $G$ is antisymmetric on $\sigma \leftrightarrow \be$. Then it vanishes when contracted with the symmetric factor $\dot x^\sigma \dot x^\be$ of (\ref{28}). We are finally left with
\begin{eqnarray}\label{31}
\ddot x^\al + \g^\al{}_{\be \ga}\dot x^\be \dot x^\ga = 0,
\end{eqnarray}
and $\g$ is given by
\begin{eqnarray}\label{32}
\g^\al{}_{\be \ga} = \frac{x^\al}{a^2}g_{\be \ga},
\end{eqnarray}

\ni where (\ref{31}) is the equation of a geodesic line: the particle chooses the trajectory with the shortest length. Moreover, the principle of least action gave us the Christoffel symbol or affine connection $\g^\al{}_{\be \ga}$. Once again, the ``static'' concepts of differential geometry (geodesic line and second fundamental form $\g$) were discovered via a dynamical realization. In the limit $a \rightarrow + \infty$, the equation of motion tends to 
\begin{eqnarray}\label{33}
\ddot x^\al = 0,
\end{eqnarray} 
which corresponds to the motion of a free particle (in flat bi-dimensional space) since $\g^\al{}_{\be \ga} \rightarrow 0$, in accordance with our intuition. 

In the next Section we will solve the equations of motion (\ref{31}). It will be accomplished  in two different ways.   The first one is by exploring the geometric setup that the model was constructed and the second one is by using the conserved currents obtained from Noether theorem \cite{no}.

\section{Solution to equations of motion}

Let us now obtain the solution of the equations of motion (\ref{31}) in the commutative plane. It will be obtained via two different approaches. In the first one, we will use the geometric structure of the problem, i.e., since the particle is free, it is supposed to describe a circumference of radius $a$ with constant angular velocity. Besides, we will also use the Noether theorem which provides two integrals of motion, which allow us to find the general solution of equations of motion.   

\subsection{Solving equations of motion: geometrical point of view}

There is a standard way to solve the equations of motion in different models: if we know a particular solution, the general one is obtained by applying a group transformation in which the model is based on. For example, in \cite{ry}, the author finds general spinors connected with an arbitrary state of motion of the Dirac electron by boosting plane wave solutions of the Dirac equation for a particle at rest. We will use the same prescription here. Initially, we take the following particular solution,
\begin{eqnarray}\label{34}
x^i(t) = \left( \begin{array}{cccc} 
0 \\ 
a\sin \om t \\
a \cos \om t
\end{array}
\right),
\end{eqnarray}   

\ni that describes our free particle with constant (and arbitrary) angular velocity $\om$ constrained to the 2-sphere of radius $a$. A direct calculation shows that it satisfies (\ref{31}). We have restricted the motion to the plane $x^2 x^3$.  The general solution is achieved by three successive passive rotations around $x^1$, $x^2$ and $x^3$ axes.  The rotations introduce three new and arbitrary parameters which, combined with $\om$, complete the necessary number of four constants of integration concerning the second order equation (\ref{31}). Denoting $\mathcal R_{x^i}(\ta_j)$ the rotation around $x^i$-axis by an angle $\ta_j$, we have
\begin{eqnarray}\label{35}
x^i(t) = \left[\mathcal R_{x^3}(\ta_3) \right]^i{}_j \left[\mathcal R_{x^2}(\ta_2) \right]^j{}_k \left[\mathcal R_{x^1}(\ta_1) \right]^k{}_l y^l(t),
\end{eqnarray} 
where, for example,
\begin{eqnarray}\label{36}
\mathcal R_{x^1}(\ta_1) = \left(
\begin{array}{cccc}
1 & 0 & 0 \\
0 & \cos \ta_1 & \sin \ta_1 \\
0 & -\sin \ta_1 & \cos \ta_1
\end{array}
\right). 
\end{eqnarray}
The other matrices $\mathcal R_{x^2}(\ta_2)$ and $\mathcal R_{x^3}(\ta_3)$ are well-known from the $SO(3)$-group. The parameters $\ta_i$ are the Euler angles, taken in the $x^1 x^2 x^3$ convention. For different representations of the Euler angles, see for example, \cite{gold, saku}. 

So, for the general solution one can obtain that
\begin{eqnarray}\label{37}
x^i(t) = \left( 
\begin{array}{ccc}
a\sin\ta_2\cos\ta_3\cos(\om t + \ta_1)+ a\sin\ta_3\sin(\om t+\ta_1) \\
-a\sin\ta_2\sin\ta_3\cos(\om t + \ta_1)+ a\cos\ta_3\sin(\om t+\ta_1) \\
a\cos\ta_2\cos(\om t + \ta_1)
\end{array}
\right).
\end{eqnarray}
In Section 3, we have withdrawn the variable $x^3$ from the description. One may check that the expression above obeys the identity,
\begin{eqnarray}
x^3(t) \equiv \sqrt{a^2 - (x^1(t))^2 - (x^2(t))^2}.
\end{eqnarray} 
Then, the physical solution is given by the projection of $x^i=x^i(t)$ onto the plane $x^1 x^2$. On this plane, the trajectory is an ellipse. In fact, with no loss of generality\footnote{The only effect of the last rotation $\mathcal R_{x^3}(\ta_3)$ is to make the semi-axes of the ellipse not coincident with the coordinate axes $x^1$ and $x^2$. Thus, for simplicity, we obtain the trajectory by looking to the solution $\tilde x^\al$ in (\ref{39}).} we take to the solution
\begin{eqnarray}\label{38}
\tilde x^i(t) = \left[\mathcal R_{x^2}(\ta_2) \right]^i{}_k \left[\mathcal R_{x^1}(\ta_1) \right]^k{}_l y^l(t)
\end{eqnarray} 
in the plane $x^1x^2$,
\begin{eqnarray}\label{39}
\tilde x^\al(t) = \left(
\begin{array}{ccc}
a\sin\ta_2\cos(\om t + \ta_1) \\
a\sin(\om t+ \ta_1)
\end{array}
\right).
\end{eqnarray}
The trajectory is obtained by excluding the time of the parametric equations (\ref{39}). It is given by
\begin{eqnarray}\label{40}
\frac{(\tilde x^1)^2}{a^2 \sin^2\ta_2}+\frac{(\tilde x^2)^2}{a^2} = 1,
\end{eqnarray}
which is the equation of an ellipse.

Finally, the general solution that we were looking for is given by the projection of (\ref{37}) in the plane $x^1x^2$, 
\begin{eqnarray}\label{41}
x^\al(t) = \left( 
\begin{array}{ccc}
a\sin\ta_2\cos\ta_3\cos(\om t + \ta_1)+ a\sin\ta_3\sin(\om t+\ta_1) \\
-a\sin\ta_2\sin\ta_3\cos(\om t + \ta_1)+ a\cos\ta_3\sin(\om t+\ta_1) 
\end{array}
\right),
\end{eqnarray}
whose trajectory is an ellipse. One then can ask about the possibility of interpreting this movement as generated by a central field. It will be discussed in Section 5. Our next step consists of finding $x^\al=x^\al(t)$ with the help of conserved quantities.  
 
\subsection{Solving equations of motion: conserved quantities}

One of the most impressive results in classical mechanics is the Noether theorem: if an action is invariant under a global transformation, then there is  a related integral of motion, known as Noether charge. In our case, we may look at (\ref{22}) or (\ref{25}) since they are equivalent. Considering that (\ref{22}) has global $SO(3)$-invariance,
\begin{eqnarray}\label{42}
x^i \rightarrow x'^i = R^i{}_jx^j; \, \mbox{where} \,\, R^T = R^{-1}.
\end{eqnarray} 
It implies the conservation of angular momentum,
\begin{eqnarray}\label{43}
L_i = m \varepsilon_{ijk}x^j \dot x^k \Rightarrow \frac{dL_i}{dt}=0.
\end{eqnarray}
One may also look at the expression (\ref{25}), which is invariant under time translations
\begin{eqnarray}\label{44}
t \rightarrow t' = t + \tau.
\end{eqnarray}
In this case, the corresponding conserved quantity is
\begin{eqnarray}\label{45}
E = \frac m2 g_{\al \be}(x) \dot x^\al \dot x^\be.
\end{eqnarray}

\ni where $E$ is considered as the energy of the particle. We now turn our attention to the equation of motion (\ref{31}). It is immediately decoupled if we use (\ref{45}),
\begin{eqnarray}\label{46}
\ddot x^\al + \frac{x^\al}{a^2}g_{\be \ga} \dot x^\be \dot x^\ga = 0\Rightarrow \ddot x^\al + \frac{2E}{ma^2}x^\al=0.
\end{eqnarray} 
Thus, the solution of (\ref{46}) can promptly be written
\begin{eqnarray}\label{47}
x^\al(t) = A^\al\sin(\Omega t + \varphi_\al); \, \Omega = \frac{\sqrt{2mE}}{ma}
\end{eqnarray}
where $A^\al$ and $\varphi_\al$ are arbitrary constants of integration. Substitution of solution (\ref{47}) in (\ref{43}) and (\ref{45}) gives, respectively,
\begin{eqnarray}\label{48}
\frac{L_3}{m \Omega}=- A^1 A^2 \sin(\varphi_2 - \varphi_1), \\ \label{49}
(A^1)^2+(A^2)^2 = a^2 + \frac{L^2_3}{2mE}.
\end{eqnarray} 

\ni And (\ref{48}) means that the angle between $x^1(t)$ and $x^2(t)$ is $\varphi_2 - \varphi_1$. If we assume that $\varphi_2-\varphi_1 = \frac\pi 2$, then the  general solution may be achieved by rotating the particular solution with this restriction. So, first if we substitute (\ref{48}) in (\ref{49}) we have that
\begin{eqnarray}\label{50}
(A^1)^2 + (A^2)^2 = a^2 + \frac{(A^1)^2(A^2)^2}{a^2} \Rightarrow A^1 = a \Rightarrow A^2=-\frac{L_3}{\sqrt{2mE}}.
\end{eqnarray}
We have then a particular solution $x^\al_p = x^\al_p(t)$, where $x^1_p$ and $x^2_p$ are perpendicular,
\begin{eqnarray}\label{51}
x^\al_p(t)=\left(
\begin{array}{ccc}
a\sin(\Omega t+\varphi_1) \\
-\frac{L_3}{\sqrt{2mE}}\cos(\Omega t + \varphi_1)
\end{array}
\right).
\end{eqnarray}

\ni A final general solution can be obtained by rotating the particular solution above in an active way,
\begin{eqnarray}\label{52}
\left(\begin{array}{cccc}
x^1 \\
x^2
\end{array}
\right) = \left(\begin{array}{ccc}
\cos\varphi_2 & \sin\varphi_2 \\
-\sin\varphi_2 & \cos\varphi_2
\end{array}
\right) \left(\begin{array}{cccc}
x^1_p \\
x^2_p
\end{array} \right),
\end{eqnarray}
that is,
\begin{eqnarray}\label{53}
x^\al(t) = \left(\begin{array}{cccc}
a\cos\varphi_2\sin(\Omega t+\varphi_1)-\frac{L_3}{\sqrt{2mE}}\sin\varphi_2\cos(\Omega t+\varphi_1) \\
-a\sin\varphi_2\sin(\Omega t+\varphi_1)-\frac{L_3}{\sqrt{2mE}}\cos\varphi_2\cos(\Omega t+\varphi_1)
\end{array} 
\right).
\end{eqnarray}
As expected, we have four constants of integration: $\varphi_{1,2}$, $E$ and $L_3$. Equivalence between the two solutions (\ref{41}) and (\ref{53}) is manifest if we write
\begin{eqnarray}\label{54}
\begin{array}{cccc}
\om = \Omega, \\
\ta_1 = \varphi_1, \\
\ta_3 = \varphi_2 + \frac{\pi}{2}, \\
a\sin\ta_2 = \frac{L_3}{\sqrt{2mE}}.
\end{array}
\end{eqnarray}

In the next Section, we will discuss a possible interpretation of the solution of the equations of motion in terms of an effective central potential induced by the space curvature. 

\section{Equivalence between a central force problem and the particle over a 2-sphere}

The movement of the particle over the 2-sphere was completely described so far by the physical variables $x^\al(t)$, $\al=1,2$, see (\ref{41}) or (\ref{53}). Since the trajectory is an ellipse, one may think that it could be derived by a central field. So, the objective of this section is to show that the solution $x^\al(t)$ is equivalent to the one described by an isotropic harmonic oscillator. We already know the time evolution of the particle. 
The idea is, instead of solving a differential equation of motion, we would like to obtain it. For that, we will use polar coordinates $(x^1,x^2)\leftrightarrow (r, \ta)$
\begin{eqnarray}\label{55}
\begin{array}{ccc}
x^1 = r \cos \ta \\
x^2 = r \sin \ta
\end{array}
\Leftrightarrow
\begin{array}{cccc}
r = \sqrt{(x^1)^2 + (x^2)^2} \\
\ta = \arctan \left(\frac{x^2}{x^1} \right).
\end{array}
\end{eqnarray} 
For simplicity, we have used the solution (\ref{39}). Let us construct the differential equations obeyed by the coordinates $\ta$ and $r$. We have that
\begin{eqnarray}\label{a1}
\ta(t) = \arctan \left(\frac{x^2(t)}{x^1(t)} \right) = \arctan \left(\frac{\sin\Delta}{\sin \ta_2 \cos\Delta} \right), \\ \label{a2}
r(t) =a \sqrt{\sin^2\ta_2\cos^2\Delta + \sin^2\Delta}, \qquad \qquad
\end{eqnarray}
where we have used the shorthand notation $\Delta = \om t+ \ta_1$. First time derivative of (\ref{a1}) gives
\begin{eqnarray}\label{a3}
\dot \ta (t) = \frac{\om a^2 \sin \ta_2}{r^2(t)} = \frac{L_3}{mr^2(t)}, 
\end{eqnarray}
since the angular momentum $L_3$ is given by
\begin{eqnarray}\label{a4}
L_3 = m(\dot x^2 x^1 - \dot x^1 x^2) = m \om a^2 \sin \ta_2.
\end{eqnarray}

We turn our attention to the radial variable. It is a tedious but rather direct calculation to obtain the second order time derivative of Eq. (\ref{a2}). We have
\begin{eqnarray}\label{a5}
\dot r(t) = \frac{\om a^2 \sin 2 \Delta (1-\sin^2\ta_2)}{2r}.
\end{eqnarray}

\ni The second time derivative reads
\begin{eqnarray}\label{a6}
\ddot r = \frac{\om^2 a^2 \cos 2 \Delta (1-\sin^2\ta_2)}{r} -\frac{\om a^2 \sin 2\Delta(1-\sin \ta_2)}{2r^2}\dot r.
\end{eqnarray}

\ni Substituting $\dot r(t)$ into the expression above, one finds after rearranging the terms,
\begin{eqnarray}\label{a6.1}
\ddot r = \frac{\om^2 a^4}{r^3}[-(\cos^2 \Delta \sin^2 \ta_2 + \sin^2 \Delta)^2+
\sin^2 \ta_2(\sin^2\Delta+\cos^2\Delta)^2],
\end{eqnarray}
which multiplied by the mass of the particle becomes
\begin{eqnarray}\label{a7}
m\ddot r= -\om^2 r + \frac{L^2_3}{m^2 r^3} \Rightarrow m\ddot r = -m \om^2 r + \frac{L^2_3}{mr^3}.
\end{eqnarray}
Eqs. (\ref{a3}) and (\ref{a7}) are exactly the ones obeyed by a particle in a central field \cite{gold}. Eq. (\ref{a7}) may be seen as the second Newton's law for a particle in a isotropic harmonic oscillator. The term $\frac{L^2_3}{m r^3}$ corresponds to the centrifugal force always present when one writes a central force in polar coordinates. The first term, that has been associated with the harmonic oscillator, may be considered as an effective force due to the curved space the particle is constrained to. In fact, we construct the scalar or total curvature of the surface
\begin{eqnarray}\label{a8}
R = g^{\al \be}(\p_\ga \g^\ga{}_{\al \be} - \p_\be \g^\ga{}_{\al \ga} + \g^\ga{}_{\al \be}\g^\lambda{}_{\lambda \ga} - \g^\ga{}_{\al \lambda} \g^\lambda{}_{\be \ga}).
\end{eqnarray} 
Using the Christoffel symbols (\ref{32}) and the inverse of the metric
\begin{eqnarray}\label{a9}
g^{\al \be} = \delta^{\al \be} + \frac{x^\al x^\be}{a^2},
\end{eqnarray}
one obtains
\begin{eqnarray}\label{a10}
R = \frac{2}{a^2}.
\end{eqnarray}
It turns out that the constant force of second Newton's law (\ref{a7}) is proportional to the total curvature,
\begin{eqnarray}\label{a11}
k = m \om^2 = m\frac{2mE}{m^2 a^2} = RE. 
\end{eqnarray}
Thus the movement of the free particle over a 2-sphere projected in $x^1x^2$-plane is equivalent to the movement described by a particle in a central effective potential
\begin{eqnarray}\label{60}
V_{eff}(r) = \frac{RE}{2}r^2 + \frac{L_3}{2m r^2},
\end{eqnarray}
as stated and both potentials, $V(r)\sim \frac1r$ and $V(r)\sim r^2$ produce the same trajectory, i.e., an ellipse. 



\section{Noncommutative classical mechanics in a curved phase-space}

In this Section we will assume a symplectic structure for the classical mechanics of a particle in a curved phase-space.  The target geometry is the 2-sphere described above.  We will demonstrate that there is a correction to Newton's second law thanks to the curved configuration of the phase-space, which shows that the space configuration alone can bring consequences to the result. Namely, we will see that in flat space, what causes a NC correction in the potential function. In the 2-sphere curved space we will see that there is a NC correction without the existence of a potential effect over the particle.

Let us begin by describing the appearance of the NC contribution in generalized (without a specific potential) Newton's second law \cite{rsv}.  We can define a theory as being formulated by a set of canonical variables $\xi^a$, where $a=1,\ldots,2n$ combined with a symplectic structure $\{\xi^a, \xi^b\}$.   This structure can be extended in order to accommodate arbitrary function of $\xi^a$ such as

\bee
\label{A}
\{F,G\}\,=\,\{\xi^a , \xi^b \} \frac{\p F}{\p \xi^a}\frac{\p G}{\xi^b}
\eee

\ni which can be used, of course, in classical mechanical systems as the one we will analyze in this work.

In Hamiltonian systems, we can use the structure given in (\ref{A}) to write the equations of motion for a Hamiltonian given by $H=H(\xi^a)$ such that

\bee
\label{B}
\dot{\xi^a} = \{\xi^a , H\}
\eee

\ni and for a generalized function $F$ defined in this space we can write that

\bee
\label{C}
\dot{F} = \{F,H\}.
\eee

\ni In our case, we will consider a phase-space given by $\xi^a=(x^i, p_i),\;i=1,2,3$.   The algebra between these coordinates are

\bee
\label{D}
\{ x^i , x^j\}\,=\,\theta^{ij}, \quad  \quad \{ x^i , p_j\}\,=\, \delta^i_j, \quad  \quad    \{ p_i , p_j\}\,=\,0, \quad 
\eee

\ni where $\hbar\,\theta^{ij}$ must have dimension of area.  Let us assume that this so-called NC parameter is of the Planck's area order, i.e., $l^2_P = \hbar G/c^3$, so we have that the tensor $\theta^{ij}$ must be of $G/c^3$ order.   Hence, in the classical limit, the symplectic framework will not have $\hbar$ \cite{rsv}.   This result agrees with this kind of limit.

Let us consider two arbitrary functions $F$ and $G$, defined on the phase-space.  Using Eqs. (\ref{A}) and (\ref{D}) we have that

\bee
\label{E}
\{F,G\}\,=\,\theta^{ij}\,\frac{\p F}{\p x^i}\frac{\p G}{x^j}\,+\,\frac{\p F}{\p x^i}\frac{\p G}{p_i}\,-\,\frac{\p F}{\p p_i}\frac{\p G}{x^i}
\eee

\ni  and for example, if we have a Hamiltonian of the standard form

\bee
\label{F}
H\,=\,\frac{p_i p^i}{2m}\,+\,V(x)
\eee

\ni using (\ref{C}) and (\ref{E}) we have the equations of motion given by

\baa
\label{G}
\dot{x}^i \,&=&\,\frac{p^i}{m}\,+\,\theta^{ij}\frac{\p V}{\p x^j} \nonumber \\
\dot{p}^i \,&=&\,-\,\frac{\p V}{\p x_i}\,\,.
\eaa

\ni   Notice from these both equations that an obvious conclusion is that if $V=0$ we have $p_i=$constant and $x^i$ is a linear function of time.
Newton's second law can be obtained analogously and the result is

\bee
\label{H}
\ddot{x}^i\,=\,-\,\frac{\p V}{\p x_i}\,+\,m \theta^{ij}\,\frac{\p^2 V}{\p x^j \p x_k} \dot{x}_k \,\,.
\eee

\ni This result was used to investigate several models in physics \cite{nossojhep}.   Here, we want to verify how the phase-space curvature affects the NC contribution.

In our case, we want to discuss the NC approach for the free particle in a flat 3D space which has the Lagrangian given by

\bee
\label{I}
L_{ph}\,=\,\frac m2 g_{\alpha\beta} \dot{x}^\alpha \dot{x}^\beta
\eee

\ni where $g_{\alpha\beta}$ is given in (\ref{26}).   From  (\ref{I}) we have that

\baa
\label{J}
p_x &=& m\dot{x} \,+\, \frac{mx(x\dot{x}+y \dot{y} )}{a^2 - x^2 -y^2} \nonumber \\
p_y &=& m\dot{y} \,+\, \frac{my(x\dot{x}+y \dot{y} )}{a^2 - x^2 -y^2}
\eaa

\ni where we have used that $x_1=x$ and $x_2=y$.   From Eqs. (\ref{I}) and (\ref{J}), the Hamiltonian is given by

\bee
\label{K}
H\,=\,\frac{1}{2ma^2} \bigglb[ (a^2 -x^2) p^2_x\,+\,(a^2 - y^2 ) p^2_y\,-\,2p_x p_y xy \biggrb]
\eee

\ni  and our set of symplectic variables is given by $\xi =(x,y,p_x ,p_y )$. Using Eqs. (\ref{C}) and (\ref{E}) and the Hamiltonian in (\ref{K}) we have the NC equations of motion

\baa
\label{K1}
\dot{x} &=& \frac{1}{m a^2} \bigglb[ (a^2 -x^2) p_x\,-\,xyp_y \,-\,\theta \biglb( y p^2_y\,+\,xp_x p_y \bigrb) \biggrb] \nonumber \\
\dot{y} &=& \frac{1}{m a^2} \bigglb[ (a^2 -y^2) p_x\,-\,xyp_x \,-\,\theta \biglb( x p^2_x\,+\,yp_x p_y \bigrb) \biggrb]  \nonumber \\
\dot{p_x} &=&  \frac{1}{ma^2} \biglb( x p^2_x \,+\, p_x p_y y \bigrb) \\
\dot{p_y} &=& \frac{1}{ma^2} \biglb( y p^2_y \,+\, p_x p_y x \bigrb) \,\,. \nonumber 
\eaa

\ni  Notice that when $\theta=0$ we have the standard commutative phase-space equations of motion.   Secondly, from (76) we can see the effect of a curved phase-space.  For a free particle we must have $\dot{p}_x = \dot{p}_y =0$, since this is the result of a free particle in a flat phase-space.  But it is an easy obtainable conclusion.  However, before the calculation of $\dot{p}_x$ or $\dot{p}_y$ we can see the curvature effect already in $\dot{x}$ and $\dot{y}$.   In other words, we do not need the values of $\dot{p}_x$ and $\dot{p}_y$ to know that the curvature plays a kind of potential in order to perturb the NC calculations.  In this way we can ask if, at Planck's scale, we have only the curved spaces and, if it does mean that we have naturally quantum features and gravity.  Since both curvature and gravity are connected, we could now add another ingredient and say that quantization curvature and gravity are intrinsically connected through noncommutativity.   In other words, a curved space can have hidden quantum features that are disclosed since its space coordinates obey a NC Poisson bracket algebra since the equations of motion are NC.

After a long algebra the NC Newton's second law for our particle on the 2-sphere is

\bee
\label{K2}
m \ddot{x} \,=\, -\,\frac{1}{m a^4} \bigglb[ x(a^2 - x^2 ) p_x^2\,-\, 2 x^2 y p_x p_y - x(a^2 + y^2 ) p_y^2 \biggrb] \,-\,\frac{\theta}{m a^2} (p_x^2 \,+\, p^2_y ) p_y
\eee

\ni and

\bee
\label{K3}
m \ddot{y} \,=\, -\,\frac{1}{m a^4} \bigglb[ y(a^2 - y^2 ) p_y^2\,-\, 2 x y^2 p_x p_y - y(a^2 + x^2 ) p_x^2 \biggrb] \,-\,\frac{\theta}{m a^2} (p_x^2 \,+\, p^2_y ) p_x
\eee

\ni and curiously we can notice that in  (\ref{H}) the NC correction depends on the background space through the $\theta^{ij}$ parameter and also on the variations of the potential.   This result could lead us to think that for our free particle, the NC corrections would be zero, as the expression obtained in \cite{rsv} (Eq. (\ref{H})) could indicate).  However, we can see that the curvature of the space originates a NC correction as well, in spite of a zero potential.

With these results in hand, it is natural to investigate the divergences in the case of a quantization.  This will be accomplished in the next section.

\section{Coherent states quantization}

In 1926, Schr\"{o}dinger introduced the concept of coherent states in physics.   The motivation was the harmonic motion of a particle in quantum mechanics.  He demonstrated that if the particle is in a coherent state, then its motion is analogous to the motion of the corresponding classical particle.  The coherent states were used in potentials different from the harmonic oscillator.  In the discussion on interacting bosons and in a field theory concerning a fermionic coherent state \cite{mrnuy}.   The main motivation to investigate coherent states are the facts that they can describe the quantum state of a laser and that they can describe superfluids and superconductors.

In the quantization process in a NC space we have to begin by promoting the NC coordinates $x$ and $y$ to operators in Hilbert space as

\bee
\label{aa1}
\Big[ \hat{x}, \hat{y} \Big]\,=\,i\theta  \quad , \quad \Big[ \hat{x}, \hat{p}_x \Big]=\Big[ \hat{y}, \hat{p}_y \Big]= i \quad , \quad \Big[ \hat{p}_x, \hat{p}_y \Big]\,=\,0
\eee

\ni  where we have used that $\hbar=c=1$.   As in the last section, $\theta$ has length squared dimension.  It is also a measure of the noncommutativity of the coordinates.  And we cannot speak in terms of points since the space has became blurry.

Let us introduce a proper set of states and a set of operators such as

\baa
\label{aa2}
\hat{Z} &=& \frac{1}{\sqrt{2}} ( \hat{x}\,+\,i\hat{y} ) \nonumber \\
\hat{Z}^\dagger &=& \frac{1}{\sqrt{2}} ( \hat{x}\,-\,i\hat{y} )
\eaa

\ni and both obey

\bee
\label{aa3}
\Big[ \hat{Z}, \hat{Z}^\dagger \Big]\,=\,\theta
\eee

\ni which is the Heisenberg-Fock algebra \cite{gs}.

If we substitute $\theta$ by $\hbar$, we can associate $\hat{Z}$ and $\hat{Z}^\dagger$ with the creation and annihilation operators of the standard quantum mechanics.  It is easy to realize that the classical limit can, in this way, be given by the limit $\theta \longrightarrow 0$, which is equivalent to $\hbar \longrightarrow 0$ in quantum mechanics.  Hence, from \cite{glauber}, we know that there are coherent states that are eigenstates of annihilation operator.  Now, we can construct eigenstates such as

\baa
\label{aa4}
\hat{Z} |Z> &=& z |Z> \nonumber \\
<Z| \hat{Z}^\dagger &=& <Z| \bar{z}
\eaa

\ni which have complex eigenvalues z.

The annihilation and creation operators can also be written as functions of the NC parameter $\theta$.  So, the boson Fock space, constructed as 
${\cal F}_c = \mbox{span}\{|n>,\,n\,\in\,N\,\}$, with

\bee
\label{aa5}
|N> \,=\, \frac{1}{\sqrt{n!}}\,(\hat{Z}^\dagger)^n\,|0>
\eee

\ni is isomorphic to a Hilbert space ${\cal H}$, which is the result of a NC configuration space ${\cal H}_c$ \cite{gs}.   At the quantum level, the wavefunction obeys the representation

\bee
\label{aa6}
\hat{X} \psi = \hat{x} \psi, \quad  \quad \hat{Y} \psi = \hat{y} \psi, \quad  \quad  \hat{P}_x \psi = \frac \hbar\theta \Big[ \hat{y}, \psi \Big], \quad  \quad 
\hat{P}_y \psi = -\frac \hbar\theta \Big[ \hat{x}, \psi \Big]
\eee

\ni which defines a unitary representation since in (\ref{aa6}) we have self-adjoint properties relative to the quantum Hilbert space inner product.

It is well known that normalized coherent states in the classical Hilbert space, are given by

\bee
\label{aa7}
|Z> \,=\, exp \Big( \frac{-|z|^2}{2\theta} \Big) \, exp \Big( \frac{-z \hat{Z}^\dagger}{\theta} \Big) |0>
\eee

\ni where $|z|^2 = z \bar{z}$ and the vacuum state $|0>$ is annihilated by $\hat{Z}$.  The wavefunction of a free particle on the NC plane is given by \cite{ss}

\bee
\label{aa8}
\psi_{\roarrow{p}}\,(\roarrow{x}) \equiv\,exp \Biglb( - \theta \frac{\roarrow{p}^2}{4}\,+\,i\roarrow{p}\cdot \roarrow{x} \Bigrb)\,=\,<\roarrow{p} | \roarrow{x} >_{\theta}
\eee 

\ni and the amplitude between two states of distinct mean positions can provide the Feynman path integral

\baa
\label{aax}
<\roarrow{y} | \roarrow{x} >\,&=&\, \int \frac{d^2p}{(2\pi)^2} <\roarrow{y} | \roarrow{p} ><\roarrow{p} | \roarrow{x} > \nonumber \\
&=& \frac{1}{2\pi \theta}\, exp \Biglb[ -\frac{(\roarrow{x} - \roarrow{y})^2}{2\theta} \Bigrb]
\eaa

\ni which shows us that the effect of noncommutativity in the scalar product of two mean positions is to change the Dirac $\delta$-function by a Gaussian one, where $\sqrt{\theta}$ is relative to the minimum length achievable in fuzzy space.

Hence, one can find the NC formulation of the path integral for the propagation kernel.  We  can start from the formulation of the path integral for the propagation kernel.  The formulation of the discretized transition amplitude between two close points \cite{fh} is given by

\bee
\label{aa9}
K_{\theta} (x-y;T) \,=\, N \int [Dx]\,[Dp] exp \Bigglb[ i\int^x_y \vec{p}\cdot d\vec{x} \,-\, \int^T_0 d\tau\Bigglb( H(\vec{p}\cdot \vec{x})\,+\,\frac{\theta}{2T}\vec{p}^{\,2} \Biggrb) \Biggrb] \,\,,
\eee 

\ni which is the propagation amplitude where the particle propagates from an initial position $y$ to a final position $x$.  The amplitude is given by a standard sum over all phase-space paths connecting $y$ to $x$ in a total time $T$.  Of course, thanks to noncommutativity, the final result is modified by the Gaussian factor.  In our case, the propagator of a particle in a 2-sphere surface will be given by

\baa
\label{aa10}
&&K_{\theta} (x-y;T)\,=\, N \int [Dp]\,\delta[\dot{\vec{p}}] \nonumber \\
&\times& exp \Bigglb\{ i [\vec{p}\cdot \vec{x}]^x_y  
- \int^T_0 \!\!\!\!\! d\tau \Bigglb[ \frac{1}{2ma^2} \Big[a^2 p_x^2 + a^2 p^2_y - x^2 p^2_x - y^2 p_y^2 -2p_xp_y xy \Big] \
\!\!\!\!+\frac{\theta}{2T} \Big(\vec{p}^{\,2}_x+\vec{p}^{\,2}_y \Big) \Biggrb] \Biggrb\} \nonumber \\
&=& \int \frac{d^3 p}{(2\pi )^2} \,e^{i\vec{p}\cdot (\vec{x}-\vec{y})}\,exp \Big\{-\,\frac{1}{2ma^2} \Big[ a^2 (T+m\theta )\vec{p}^{\;2}\,-\,T\,(\vec{p}\cdot \vec{x} )^2 \Big] \Big\} 
\eaa

\ni where we have used the Hamiltonian given in Eq. (\ref{K}).  After some calculation we have that the final form of the propagation kernel is 

\baa
\label{aa11}
&&K_{\theta} (x-y;T)\,=\,\frac{ma^2}{2\pi} \Big[ \Big( T(a^2 - x^2)\,+\,ma^2 \theta \Big)\,\Big( T(a^2 - y^2)\,+\,ma^2 \theta \Big) \Big]^{-\frac 12} \nonumber \\
&\times& exp \Bigglb[ \frac{[Tp_y xy + i(\vec{x}\cdot \vec{y} )_x ]^2 ma^2}{2[T(a^2 - x^2 )+ma^2 \theta ]} \Biggrb]
exp \Bigglb[ \frac{[Tp_x xy + i(\vec{x}\cdot \vec{y} )_y ]^2 ma^2}{2[T(a^2 - y^2 )+ma^2 \theta ]} \Biggrb]
\eaa 

\ni where $(\roarrow{x}-\roarrow{y})_x$ and $(\roarrow{x}-\roarrow{y})_y$, are the $x$ and $y$ components of $(\roarrow{x}-\roarrow{y})$ respectively.   To verify the short-time limit of the propagation kernel, we have to make $T \longrightarrow 0$ in the above $K_{\theta}(x-y;T)$.  So, the result is

\baa
\label{aaxx}
K_{\theta} (x-y;0) \,=\, \frac{1}{2\pi \theta}\, exp \Biglb[ -\frac{(\roarrow{x} - \roarrow{y})^2}{2\theta} \Bigrb]
\eaa 

\ni which agrees with the result obtained in \cite{sg2} (also for a free particle) and described in (\ref{aax}).   Eq. (\ref{aaxx}) shows, as before, that the kernel is not a $\delta$-function but a Gaussian one.  It is a consequence \cite{sg2} of the fact that the best possible place to find the particle in a NC space with curvature is also inside a cell of area $\theta$.  We can see that although the noncommutativity of the space changes the form of the propagator, the curvature of the space does not have the same effect.

Let us calculate the Green function  through the Laplace transform in the following way,

\baa
\label{aa12}
G_{\theta}(x-y;E)\,&=&\,\int_0^\infty dT\,e^{-ET}\,K_{\theta}(x-y;T) \nonumber \\
&=&\int \frac{d^3 p}{(2\pi )^2} \,e^{i\vec{p}\cdot (\vec{x}-\vec{y})}\,G_{\theta}(E;\vec{p}^{\;2})
\eaa

\ni and using Eq. (\ref{aa10}) in (\ref{aa12}) we have the Green function given by

\bee
\label{aa13}
G_{\theta}(x-y;E)\,=\,\int \frac{d^3 p}{(2\pi )^2} \,e^{i\vec{p}\cdot (\vec{x}-\vec{y})}\,
\frac{e^{-\frac{\theta \vec{p}^{\;2}}{2}}}{E + \frac{a^2\vec{p}^{\;2}-(\vec{p}\cdot \vec{x})^2}{2ma^2}}
\eee

\ni where we can see that $G_{\theta}(E;\vec{p}^{\;2})$ in Eq. (\ref{aa13}) shows also an exponential cutoff for large momenta, which arises from the noncommutativity of the coordinates, as the free particle in flat phase-space, although it has a different form from the $G_{\theta}(E;\vec{p}^{\;2})$ described in \cite{sg2}.   In $G_{\theta}(E;\vec{p}^{\;2})$ in Eq. (\ref{aa13}) we can see clearly that the curvature adds a $\:\:-\,(\vec{p}\cdot \vec{x})^2/2ma^2\:\:$ term in the denominator.   As in \cite{sg2}, the modification of the Fourier transform occurs from the mean value definition over coherent states.

In another way of quantization, it is very well known that the canonical quantization is based on Dirac brackets formalism, which will be discussed in the next section.


\section{Hamiltonization of constrained systems: interpretation of the Dirac brackets based on geometric grounds}

Since our discussion on the dynamics of a constrained system has been restricted to the Lagrangian formalism, the objective of this section is based on the hamiltonization of the Lagrangian $L_\lambda$. At the time when Dirac proposed his formalism, it was not completely understood how to introduce constraints into the Hamiltonian formalism \cite{dirac}, which is a solved problem in current days \cite{aadbook,gt,mhct,aadb}. Hamiltonization of $L_\lambda$ leads to the so-called Dirac brackets and we will provide its geometric interpretation. The construction of the Hamiltonian concerning (\ref{22}) begins with the definition of the conjugate momenta
\begin{eqnarray}\label{61}
p_A \equiv \frac{\p L}{\p \dot q^A},
\end{eqnarray}
where we wrote collectively $q^A = (x^i, \lambda)$. According to the formalism, we can use the expression of conjugate momenta to obtain the maximum number of velocities as functions of momenta and configuration variables,
\begin{eqnarray}\label{62}
p_A=\frac{\p L}{\p \dot q^A}\Leftrightarrow 
\left \{
\begin{array}{cccc}
p_i=\frac{\p L}{\p \dot x^i} \Leftrightarrow \dot x^i = \frac1m p_i, \\
p_\lambda = \frac{\p L}{\p \dot \lambda} \Rightarrow p_\lambda = 0. 
\end{array}
\right.
\end{eqnarray} 

\ni Let us define $T_1 \equiv p_\lambda = 0$ and call it as primary constraint. The complete Hamiltonian is defined in extended phase space $q^A, p_A, v$
\begin{eqnarray}\label{63}
H &=& p_A \dot q^A -L +v p_\lambda \cr
&=& \frac{1}{2m}p^2_i - \lambda[(x^i)^2 - a^2] + vp_\lambda,
\end{eqnarray}
where $v$ is a Lagrange multiplier and all velocities enter into $H$ according to (\ref{62}). 
Let us write the equations of motion via Poisson brackets again such that
\begin{eqnarray}\label{65}
\dot q^A = \{q^A,H \} \Rightarrow \left \{
\begin{array}{ccc}
\dot x^i = \frac1m p_i, \\
\dot \lambda = v,
\end{array}
\right.
\end{eqnarray}
\begin{eqnarray}\label{66}
\dot p_i = \{p_i, H\} = 2 \lambda x^i.
\end{eqnarray}
Since a constraint must be constant, one obtains the following chain of secondary constraints
\begin{eqnarray}\label{67}
T_2 = \dot p_\lambda = \{p_\lambda, H \} \Rightarrow T_2 = (x^i)^2 - a^2 = 0, \\ \label{68}
T_3 = \dot T_2 = \{T_2, H\} \Rightarrow T_3 = x^ip_i = 0, \qquad  \\ \label{69}
T_4 = \dot T_3 = \{T_3, H\} \Rightarrow T_4 = \frac1m p^2_i + 2\lambda (x^i)^2.
\end{eqnarray}
Finally, the evolution in time of $T_4$ allows us to find the Lagrange multiplier $v$,
\begin{eqnarray}\label{70}
v=0.
\end{eqnarray}
The matrix $T_{ab} = \{T_a, T_b \}$; $a,b = 1,2,3,4$ is invertible, then according to the Dirac terminology, the constraints are called second class (actually, the existence of $T^{-1}_{ab}$ is the reason why all multipliers have been found \cite{aadbook}). The Dirac brackets are
\begin{eqnarray}\label{71}
\{A,B\}^* = \{A,B\} - \{A,T_a\}T^{-1}_{ab}\{T_b,B\}.
\end{eqnarray}
So, the equations of motion are defined over the constraint surface and one may forget about the equations $T_a= 0$. They read,
\begin{eqnarray}\label{72}
\dot Y = \{ Y, H_0\}^*,
\end{eqnarray} 
where $Y=(x^i,p_i)$ since the sector $(\lambda, p_\lambda)$ may be omitted and $H_0 \equiv H -vp_\lambda$. The basic Dirac brackets for the $(x^i, p_i)$-sector have the form
\begin{eqnarray}\label{73}
\{x^i, x^j\}^* = 0, \qquad \quad \\
\{x^i, p^j\}^* = \de^{ij} - \frac{x^ix^j}{a^2}, \quad \\
\{p^i, p^j\}^* = -\frac{1}{a^2}(x^i p^j - x^jp^i). 
\end{eqnarray} 
Since the equations of motion described via Lagrangian formalism give the proper time evolution of the particle over the surface as well as the Lagrangian and Hamiltonian formulations are equivalent \cite{gt}, one expects a relation between Christoffel symbols and the Dirac bracket. To see this, first we decouple the equation for $x^i$,
\begin{eqnarray}\label{73.1}
m\dot x^i = p^i \Rightarrow m\ddot x^i = 2\lambda x^i. 
\end{eqnarray} 
With the help of the constraints $T_2$, $T_4$ and (\ref{65}), we obtain that
\begin{eqnarray}\label{73.2}
\lambda = - \frac{m(\dot x^i)^2}{2 a^2}.
\end{eqnarray}
Then,
\begin{eqnarray}\label{74}
\ddot x^i = -\frac{(\dot x^i)^2}{a^2} x^i.
\end{eqnarray} 
On the other hand, we may write
\begin{eqnarray}\label{75}
\ddot x^i = \frac{1}{m} \{p^i, H_0 \}^* |,
\end{eqnarray}
where $|$ denotes substitution of $p_j$ in terms of position and velocity variables, see (\ref{62}). The $\al$-sector ($\al = 1,2$)  of equation (\ref{74}) coincides with equations of motion (\ref{31}) of the Lagrangian formalism. Comparing it with (\ref{75}), one finds
\begin{eqnarray}\label{76}
\{H_0, p^\al \}^* | = m \g^{\al}{}_{\be \ga} \dot x^\be \dot x^\ga = -\ddot x^\al.
\end{eqnarray}
This calculation that compares equations of motion in both Lagrangian and Hamiltonian formalisms shows the intrinsic relation between Christoffel symbols and Dirac brackets, as these structures are the ones responsible for the time evolution of the particle in each formalism.   

\section{Application: spinning particle}

The complete understanding of electron spin was accomplished in the realm of quantum electrodynamics. If we consider the Dirac equation
\begin{eqnarray}\label{77}
i \hbar \p_t \Psi = \hat H \Psi; \,\, \hat H = c\al^i \hat p_i + mc^2 \be, 
\end{eqnarray}
as one-particle equation in Relativistic Quantum Mechanics then, in the Heisenberg picture, the position operators experience a quivering motion \cite{diracbook}
\begin{eqnarray}\label{78}
x^i = a^i + bp^i t + c^i\exp\{-\frac{2iH}{\hbar}t\}
\end{eqnarray}

\ni that may be considered a superposition of a rectilinear movement with an harmonic one, with high frequency $\frac{2H}{\hbar} \sim \frac{2mc^2}{\hbar}$. This harmonic oscillation was named \textit{Zitterbewegung} by Schr\"odinger \cite{sch}. In recent literature, it has been proposed a model with commuting variables that produces the Dirac equation through quantization \cite{nosso}. Analysis of the classical counterpart of the model leads to the so-called \textit{Zitterbewegung}, also experienced by spin variables. In order to provide space-time interpretation for the evolution of the classical position and spin coordinates, they were combined to produce configuration coordinates whose dynamics is given by (see details in \cite{book}),

\begin{eqnarray}\label{79}
\tilde x^i(t) &=& x^i + c\frac{p^i}{p^0}t, \\ \label{80}
J^i(t) &=& \frac{1}{2|p|}(A^i \cos \om t - B^i \sin \om t), 
\end{eqnarray}

\ni with $A^i$, $B^i$, $p^\mu$ some constants, $|p|\equiv \sqrt{-p_\mu p^\mu}$ and $\om$ has the same order of magnitude as the Compton frequency. They evolve similarly to the center-of-mass and relative position of two-body problem in a central field. The potential turns out to be $V(J) \sim J^2$; $J = |J^i|$. Assuming that (\ref{79}) and (\ref{80}) are the position variables for the electron, then $J^i$ describes an ellipse with restricted size (a particular feature of the model restricts the magnitude of $A^i$ and $B^i$ as well as their direction, since $p_iA^i = p_iB^i = 0$, center-of-mass moves perpendicularly to the plane of oscillations). According to the previous Sections, we interpret $J^i$ as the physical variables for the motion over a 2-sphere. This may explain the physical origin of the \textit{Zitterbewegung} if we assume that the electron has an internal structure \cite{raa}. It seems that Dirac himself believed that the electron was not an elementary particle, see \cite{dirac2}. The formalization of the idea developed in this Section is in progress. 

The idea of a composed electron goes back to the seminal paper by Dirac on the unitary irreducible particle representations of the Anti-de Sitter group \cite{111}. Actually, in this work, he found two remarkable representations of SO(2,3), the isometry group of Anti-de Sitter space $AdS_4$. Those representations do not have a counterpart in Poincar\'e group; they are typical of SO(2,3).  This means that, whenever the (Riemann) curvature of $AdS_4$ goes to zero, these two representations may be combined in order to construct one of the unitary irreducible representations of Poincar\'e group in terms of one-particle states. He called these representations singletons. These current days, singleton physics is an active research area \cite{112}. Moreover, preons appear as ``point-like" particles are conceived as being subcomponents of quarks and leptons. This term was coined by  Jogesh Pati and Abdus Salam in their 1974 paper \cite{114}. Preon models set out as an attempt to describe particle physics in a more fundamental level than the Standard  Model \cite{113}. In these preonic models, one postulates a set of fewer fundamental particles than those of the Standard Model, together with the interactions governing the dynamics of these fundamental particles. Based on these laws, preon models try to explain some physics beyond the Standard Model, often producing new particles and a number of phenomena which do not belong to the Standard Model.

\section{Conclusions and perspectives}

To investigate some ingredients of the formalism that can lead us to work in the Planck energy scale means to discuss the physics of the early Universe, for instance, where quantum mechanics and general relativity were combined and quantum gravity is formed.  This is one of the main motivations to study mechanisms that introduces Planck scale parameters in classical systems.  And this is one of the main motivations to use noncommutativty in order to introduce this so-called Planck scale parameter.   In this work we have analyzed the free movement of a particle upon a 2-sphere considering NC classical mechanics approach.   In this scenario, we can consider a semi-classical approach where the Planck constant was substituted by the NC parameter.   Besides, we have used the NC coherent states quantization to analyze the UV cutoff considering the divergence for large momenta.

The NC Newton's second law have shown us that the curvature of the space acted as if there was a potential since the particle flat space acceleration has the NC contribution given by the potential, namely, the NC contribution would be zero but it is not.  In the 2-sphere free particle dynamics, the NC additional term is different from  zero, which means that its origin is the curvature of the system.  This result make us to think if the curved space has quantum features hidden in its framework that can be disclosed when the coordinates obey a NC Poisson bracket algebra.  In this way we can make a connection between curvature and quantization in the same way we have a connection between mass and curvature.  

The introduction of the NC contribution make us also ask what would be the nature of the potential effect caused by the curvature.  In other words, since in a flat space free particle system the NC contribution is connected with a potential such that if $V=0$ we have no contribution, and in the curved space this effect does not happen, what is the physical meaning of this potential-type effect brought by the curvature?  And in the case of curved space and $V\not= 0$?  Where would the NC contribution appear?

On the other hand, concerning the UV divergence analysis, the final conclusion is the both flat \cite{sg2} and curved spaces have an exponential cutoff for large momenta induced by the NC parameter.   Although we have different expressions for the kernel and Green functions (comparing with flat space), the behavior is preserved.

Furthermore, we have also introduced some basic ideas of classical mechanics and differential geometry. We started by formulating the procedure of introducing constraints into the Lagrangian formalism: they were inserted via Lagrange multipliers and we have demonstrated that this procedure leads to the same number of degrees of freedom and equations of motion if we have obtained one of the variables of the known constraint and substitute it in the free Lagrangian. 
After that, we have given a detailed analysis of a particle constrained over a 2-sphere. 

Basic notions of differential geometry, such as the metric and Christoffel symbols, appear as a consequence of the description of a constrained Lagrangian system and its corresponding principle of least action.  A solution of the equations of motion was given based on geometric grounds and with the help of Noether theorem. It was also shown that physical position variables of the model evolve over an ellipse. We have proposed a central force problem whose solution for position variables are the same as those of the particle over a 2-sphere. One is led to interpret the curvature of the space where the particle lives as an origin of an effective potential. This example may be a starting point for studying general relativity.  We have also naively discussed the relation between both the Dirac brackets and Christoffel symbols, since both of them are supposed to describe the correct evolution of a particle constrained to a surface. 

Finally, as an example, we treated the so-called \textit{Zitterbewegung} of the Dirac electron. It may be seen as the effective motion of a particle over a 2-sphere, assuming that the electron bears an internal structure.       

\section{Acknowledgments}

\ni E.M.C.A. thanks CNPq (Conselho Nacional de Desenvolvimento Cient\' ifico e Tecnol\'ogico), Brazilian scientific support agency, for partial financial support.


\begin{thebibliography}{5}

\bibitem{snyder}    H. S. Snyder, Phys. Rev.  71 (1947)  38.


\bibitem{yang}  C. N. Yang, Phys. Rev.  72 (1947) 874.

\bibitem{sw} N. Seiberg and E. Witten, JHEP  9909 (1999) 032.


\bibitem{reviews}   M. R. Douglas and N. A. Nekrasov, Rev. Mod. Phys.  73 (2001) 977;
R. J. Szabo, Class. Quant. Grav.  23 (2006) R199;
R. J. Szabo, Quantum Gravity, Field Theory and Signatures of NC Space-time, arXiv:0906.2913;   R. Szabo,  Phys. Rep.  378 (2003) 207.

\bibitem{djemai}    A. E. F. Djemai, Int. J. Theor. Phys. 43 (2004) 299; A. E. F. Djemai and H. Smail, Comm. Theor. Phys. 41 (2004) 837.


\bibitem{rsv}   J. M. Romero, J. A. Santiago and J. D. Vergara, Phys. Lett. A 310 (2003) 9.





\bibitem{arno} V.I. Arnold, \textit{Mathematical Methods of Classical Mechanics}, 2nd edn. (New York: Springer, 1989).

\bibitem{aadeurop} A. A. Deriglazov, Eur. J. Phys. \textbf{29} (2008) 767-780.

\bibitem{gr} S. Weinberg, \textit{Gravitation and Cosmology: Principles and Applications of the General Theory of Relativity} (New York: John Wiley \& Sons, Inc., 1972). 

\bibitem{dirac} P. A. M. Dirac, Can. J. Math. {\bf 2} (1950) 129; \textit{Lectures on Quantum Mechanics} (New York: Yeshiva Univ., 1964).

\bibitem{sch} E. Schr\"odinger, Sitzunger. Preuss. Akad. Wiss. Phys.-Math. Kl. \textbf{24} (1930) 418. 

\bibitem{aadbook} A. A. Deriglazov, {\it Classical Mechanics, Hamiltonian and Lagrangian Formalism} (Springer-Verlag, Berlin Heidelberg, 2010).

\bibitem{manf} M. P. do Carmo, \textit{Differential Geometry of Curves and Surfaces} (New Jersey: Prentice Hall, 1976). 

\bibitem{no} E. N\"other, Nachr. d. K\"onig. Gesellsch.
d. Wiss. zu G\"ottingen, Math-phys. Klasse, (1918) 235-257. English translation by M. A. Tavel, Transport Theory and Statistical Physics, \textbf{1} (1971) 183-207.

\bibitem{ry} Lewis H. Ryder, \textit{Quantum Field Theory}
(Cambridge University Press, Cambridge, 1985).

\bibitem{gold} H. Goldstein, {\it Classical Mechanics} (Addison-Wesley, Reading, MA, 3rd. ed. 2002).

\bibitem{saku} J. J. Sakurai and Jim Napolitano, \textit{Modern Quantum Mechanics}(Addison-Wesley, San Fracisco, 2011).


\bibitem{gt} D. M. Gitman and I. V. Tyutin, \emph{Quantization of Fields with Constraints} (Berlin: Springer-Verlag, 1990).

\bibitem{mhct} M. Henneaux and C. Teitelboim, \emph{Quantization of Gauge Systems}
(Princeton: Princeton Univ. Press, 1992).

\bibitem{diracbook} P. A. M. Dirac, \emph{Principles of Quantum Mechanics} 
(Clarendon Press, Oxford, 1958).

\bibitem{nosso} A. A. Deriglazov, B. F. Rizzuti, G. P. Zamudio and P. S. Castro, J. Math. Phys. \textbf{53} (2012) 122303.

\bibitem{book} A. Deriglazov, B. Rizzuti and G. Zamudio, \textit{Spinning particles: possibility of space-time interpretation for the inner space of spin} (Lap Lambert Academic Publishing, Saarbr\"ucken, 2012).

\bibitem{dirac2} P. A. M. Dirac, Proc. Roy. Soc. Lond. \textbf{A268} (1962) 57-67.

\bibitem{111}    Paul A.M. Dirac, J. Math. Phys. 4 (1963) 901.

\bibitem{112}    M. Flato, C. Fronsdal and D. Sternheimer, arXiv: hep-th/9901043.

\bibitem{114}   J. C. Pati and  A. Salam,  Phys. Rev. D 10 (1974) 275.

\bibitem{113}   I. A. D'Souza and C. S. Kalman,  ``Preons: Models of Leptons, Quarks and Gauge Bosons as Composite Objects," World Scientific. 1992 

\bibitem{aadb} A. A. Deriglazov and B. F. Rizzuti, Phys. Rev. D {\bf 83} (2011) 125011; arXiv:1105.4171v1 [hep-th].

\bibitem{nossojhep}   E. M.C. Abreu, M. V. Marcial, Albert C. R. Mendes, Wilson Oliveira, JHEP 1311 (2013) 138, and references therein.

\bibitem{mrnuy}   A. Mann, M. Revzen, K. Nakamura, H. Umezawa and Y. Yamanaka, J. Math. Phys. 30 (1989) 2883, and references therein.

\bibitem{gs}   J. B. Geloun and F. G. Scholtz, J. Math. Phys. 50 (2009) 043505.

\bibitem{glauber}   R. J. Glauber, Phys. Rev. D 131 (1963) 2766.

\bibitem{ss}   A. Smailagic and E. Spallucci, J. Phys. A 36 (2003) L517.

\bibitem{sg2}  A. Smailagic and E. Spallucci, J. Phys. A 36 (2006) L467.

\bibitem{fh}   R. P. Feynman and A. R. Hibbs, ``Quantum mechanics and Path Integral,"  McGraw-Hill, New York, 1965.

\bibitem{raa}   B. Rizzuti, E.M.C. Abreu and P.V. Alves, ``Electron structure through a classical description of the Zitterbewegung," to appear in Phys. Rev. D (in press).


\end{thebibliography}
\end{document}